\documentclass{PoS}
\usepackage{amssymb}
\usepackage{fontenc}
\usepackage{times}
\usepackage{mathptmx}
\usepackage{graphicx}
\usepackage{natbib}
\def\hi{H\,{\sc i} }

\def\aj{AJ}%
\def\apj{ApJ}%
\def\aap{A\&A}%
\def\aapr{A\&A~Rev.}%
\title{Our changing view of the blue compact dwarf NGC 2915}

\ShortTitle{Our changing view of the blue compact dwarf NGC 2915}

\author{\speaker{E. C. Elson}, W. J G. de Blok, R. C. Kraan-Korteweg\\
       Department of Astronomy, University of Cape Town, Private Bag X3, Rondebosch 7701, South Africa\\
        E-mail: \email{elson.e.c@gmail.com}}

\abstract{Results from new \hi synthesis observations of the nearby blue compact dwarf galaxy, NGC 2915, carried out on the Australian Telescope Compact Array are presented.  High resolution \hi moment maps for the galaxy reveal complex \hi distributions and kinematics.  The presence of large non-circular velocity components within the gas at inner radii is revealed.  The central gas dynamics are consistent with the simple scenario in which winds from high-mass stars are expelling the gas outwards from the centre of the galaxy.  A model in which the central region is treated as a rotating, expanding gas torus is able to reproduce the inner \hi morphology and kinematics of NGC 2915.  We also find intriguing evidence for a gas infall event which, if confirmed, would be the first such evidence for a low-mass system.}

\FullConference{Panoramic Radio Astronomy: Wide-field 1-2 GHz research on galaxy evolution - PRA2009\\
		 June 02 - 05 2009\\
		 Groningen, the Netherlands}

\begin{document}

\section{Introduction}\label{intro}
NGC 2915 is classified as a nearby \citep[3.78 Mpc,][]{karachentsev_catalog} blue compact dwarf galaxy.  Photometry carried out by \citet{meurer1} revealed two dominant stellar populations: a compact blue stellar core which is the location of current high-mass star formation, surrounded by a diffuse, older stellar population.  \citet{meurer1} measured the average star formation rate to be $\sim$ 0.05 M$_{\odot}$ yr$^{-1}$.  Interesting is the observation that the stellar component of the galaxy is embedded in a huge ($\sim$ 22 \textit{B}-band scale lengths), low surface brightness \hi disk with well-defined spiral structure, but with apparently no significant star formation in the outer parts.  The absence of stars in the outer disk does, however, make NGC 2915 an ideal candidate for dark matter studies.  \citet{meurer2} were the first to study the dark matter distribution of a blue compact dwarf galaxy by mass-modeling the observed rotation curve of NGC 2915.  They showed that NGC 2915 is dark-matter-dominated at nearly all radii with a very dense and compact dark matter core.
 
New \hi synthesis observations of NGC 2915 using the Australian Telescope Compact Array have been carried out.  These data are used to study the gas dynamics of this galaxy in detail.  Previous investigators have focused on explaining the spiral structure of the outer disk of NGC 2915.  In this work we focus on the innermost \hi morphology of the galaxy.  In particular, an effort is made to link the energy output of the central high-mass stellar population to the central \hi morphology which, for the first time, is clearly resolved by our observations.  Details of the results will be presented by Elson et al. in prep.

\section{Data acquisition and reduction}\label{data}
NGC 2915 was observed with 6 different ATCA configurations using all 6 antennas between 23 October 2006 and 2 June 2007.  Each of the EW-352, 750D, 1.5B and 1.5C runs was approximately 12 hours long while 24 hours were spent in the 6A configuration.  Archival data obtained by \citet{meurer2} were also incorporated yielding a total of $\sim 100$ hours worth of on-source data.    

The MIRIAD software package was used to take the raw $uv$ data from the correlator through to the image analysis stage.  Standard reduction techniques were used to produce an \hi data cube.  The source data were calibrated and then continuum subtracted by fitting and subtracting a first order polynomial to the line-free channels.  The remaining $uv$ data were mapped to the image plane.  All of our data cubes were produced using natural weighting.   The dirty image was deconvolved using a Steer Clean algorithm \citep{steer_clean}.  Each of the clean components was convolved with a Gaussian approximation of the dirty beam which, at the 50$\%$ level, had a size of $17''\times  18.2''$.  The channel width of the final data cube is 3.5 km s$^{-1}$.   The noise in a line-free channel is Gaussian distributed with a standard deviation of $\sigma \sim$ 0.6 mJy beam$^{-1}$.  The channel maps will be presented in Elson et al. in prep.

\hi total intensity and velocity dispersion maps were constructed from the \hi data cube by calculating the zeroth and second moments respectively of the \hi line profiles.  Rather than constructing a conventional intensity-weighted-mean \hi velocity field from the first moments of the line profiles, third order Gauss-Hermite polynomials \citep{GHpolynomials} were fitted and the velocities associated with the fitted peaks were used to construct an \hi velocity field.

\section{\hi data products}\label{hi_products}
\subsection{Total intensity map}
The SINGS \citep{SINGS} 3.6 $\mu$m image as well as the \hi moment maps are presented in Fig. \ref{moments}.  The diffuse, old stellar population is well contained within the \hi disk.  Clearly visible in the \hi total intensity map (Fig. \ref{moments}, upper right panel) are two central \hi over-densities which our observations are able to clearly resolve for the first time.  Also noticeable is the plume-like \hi feature located towards the North-West.  From the total intensity map the total \hi mass of the galaxy is estimated to be $\sim$ 5.0$\times$10$^{8}$ M$_{\odot}$.

Several authors have attempted to explain the observed \hi morphology of NGC 2915.  \citet{masset_bureau_2003}, using numerical simulations, showed it to be unlikely that the spiral structure is caused by a rotating central bar or a rotating tri-axial DM halo.  Furthermore, NGC 2915 seems to be isolated.  The effect of a gravitational torque exerted on the disk by a nearby companion cannot be invoked.  It is therefore unlikely that the spiral wave is triggered by an external perturber.  \citet{masset_bureau_2003} also investigated the possibility that a large fraction of unseen mass is distributed in the disk of NGC 2915.  They found that a heavy disk scenario can adequately account for the observed \hi morphology but fails to describe the observed kinematics.  Finally, \citet{bureau_1999} showed that swing amplification of the spiral density wave is far too inefficient at all radii to be able to sustain the observed \hi spiral morphology.  \citet{meurer1} carried out deep imaging of the outer disk of NGC 2915 and found no significant faint stellar population.  The gas dispersion of $\sim 10$ km s$^{-1}$ of the outer disk is therefore difficult to explain in the apparent absence of stars.  

\subsection{Velocity field}
The 3rd order Gauss-Hermite \hi velocity field is displayed in the lower left panel of Fig. \ref{moments} and is clearly one of a rotating disk.  The overall ``S-shaped'' distortion of the contours suggests the presence of a kinematic warp within the disk.  Small wiggles along the outer contours are caused by streaming gas motions along the spiral arms.  The sharp kinks at inner radii are indicative of significant non-circular gas motions.  Finally, differences in the shapes of the iso-velocity contours on the receding and approaching halves of the galaxy as well as possibly non-orthogonal kinematic major and minor axes suggest a certain degree of kinematic lopsidedness.

\subsection{Velocity dispersion map}
\begin{figure}[h]
	\begin{centering}
	\includegraphics[angle=-90,width=0.5\columnwidth]{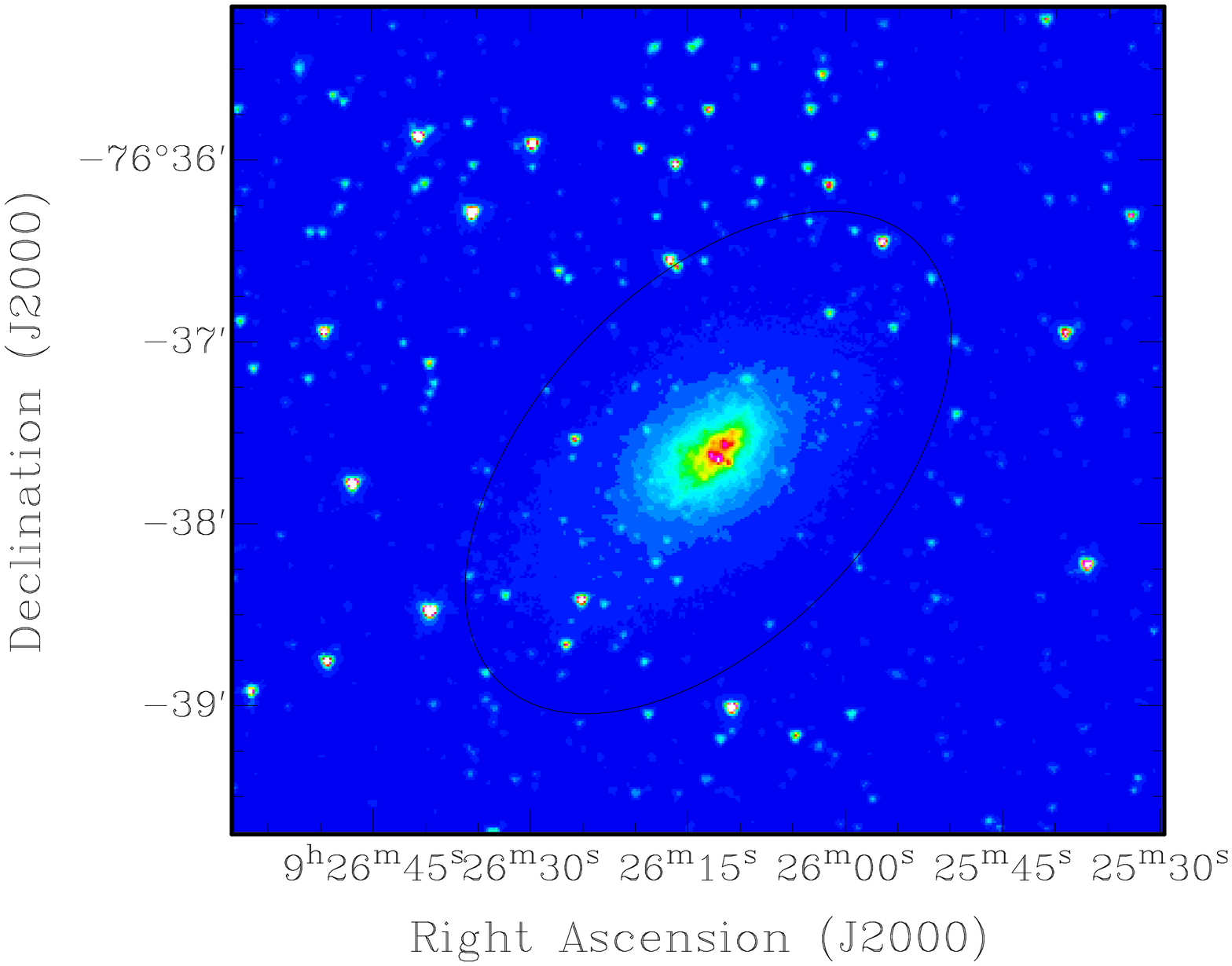}\includegraphics[angle=-90,width=0.5\columnwidth]{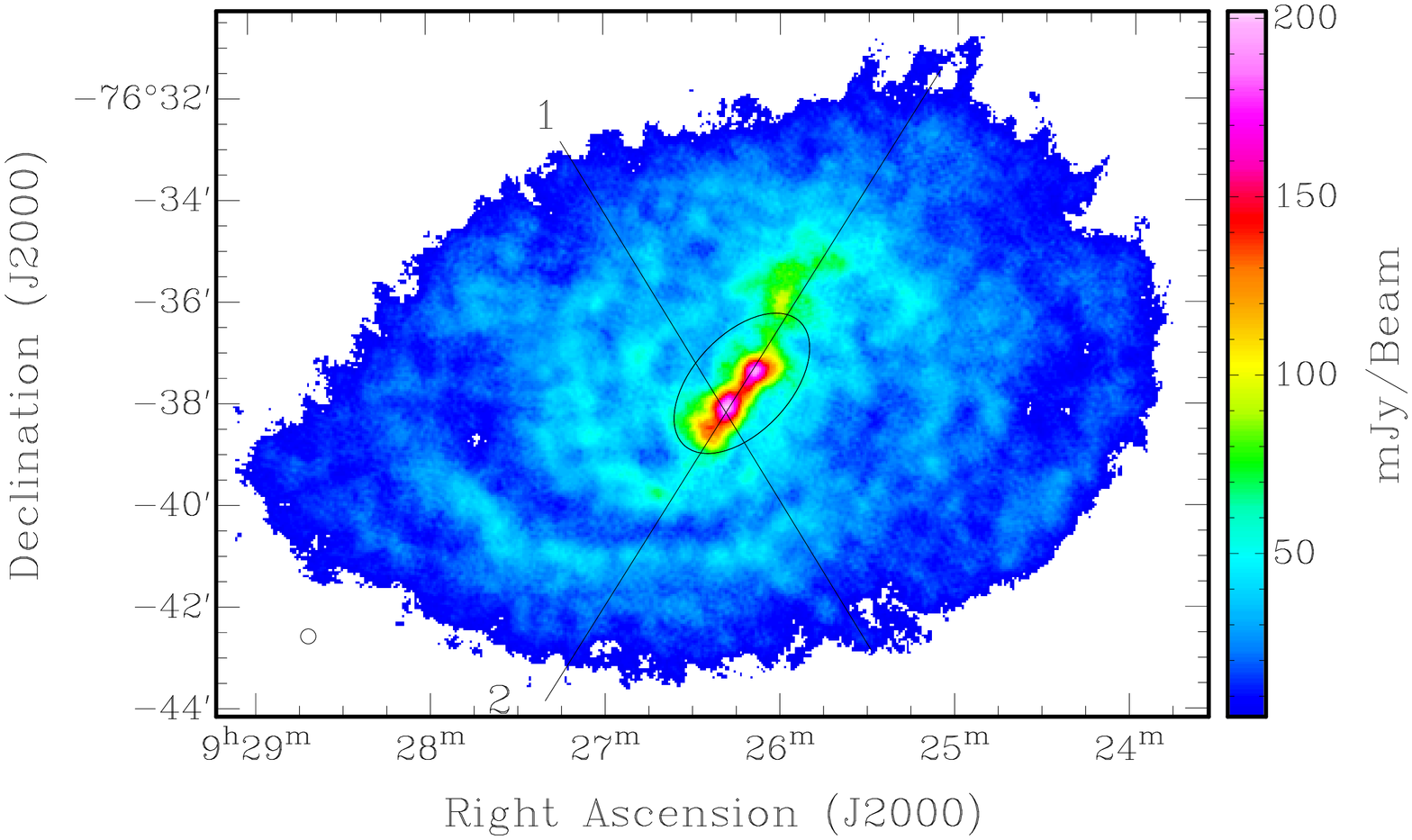}\\
	\includegraphics[angle=-90,width=0.5\columnwidth]{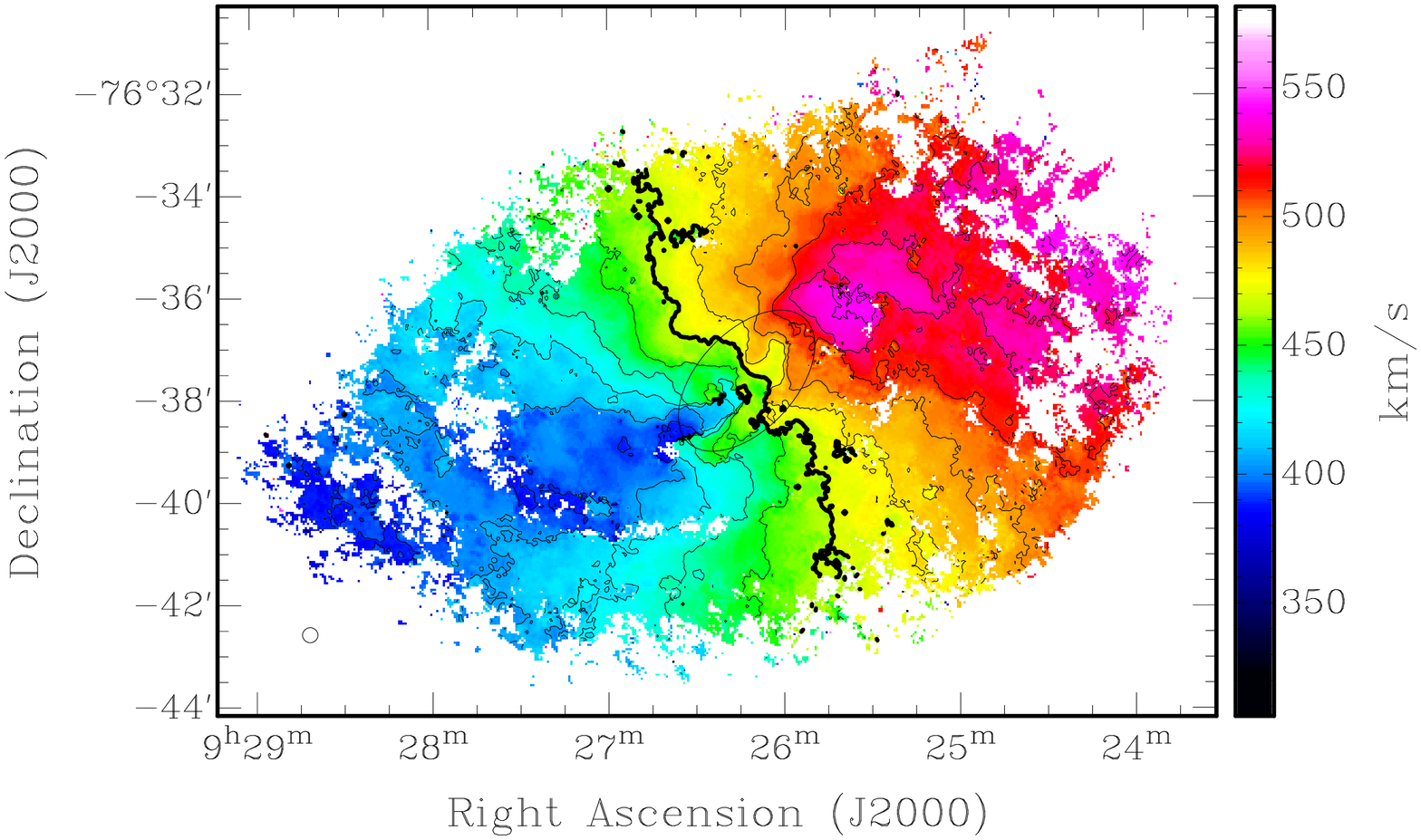}\includegraphics[angle=-90,width=0.5\columnwidth]{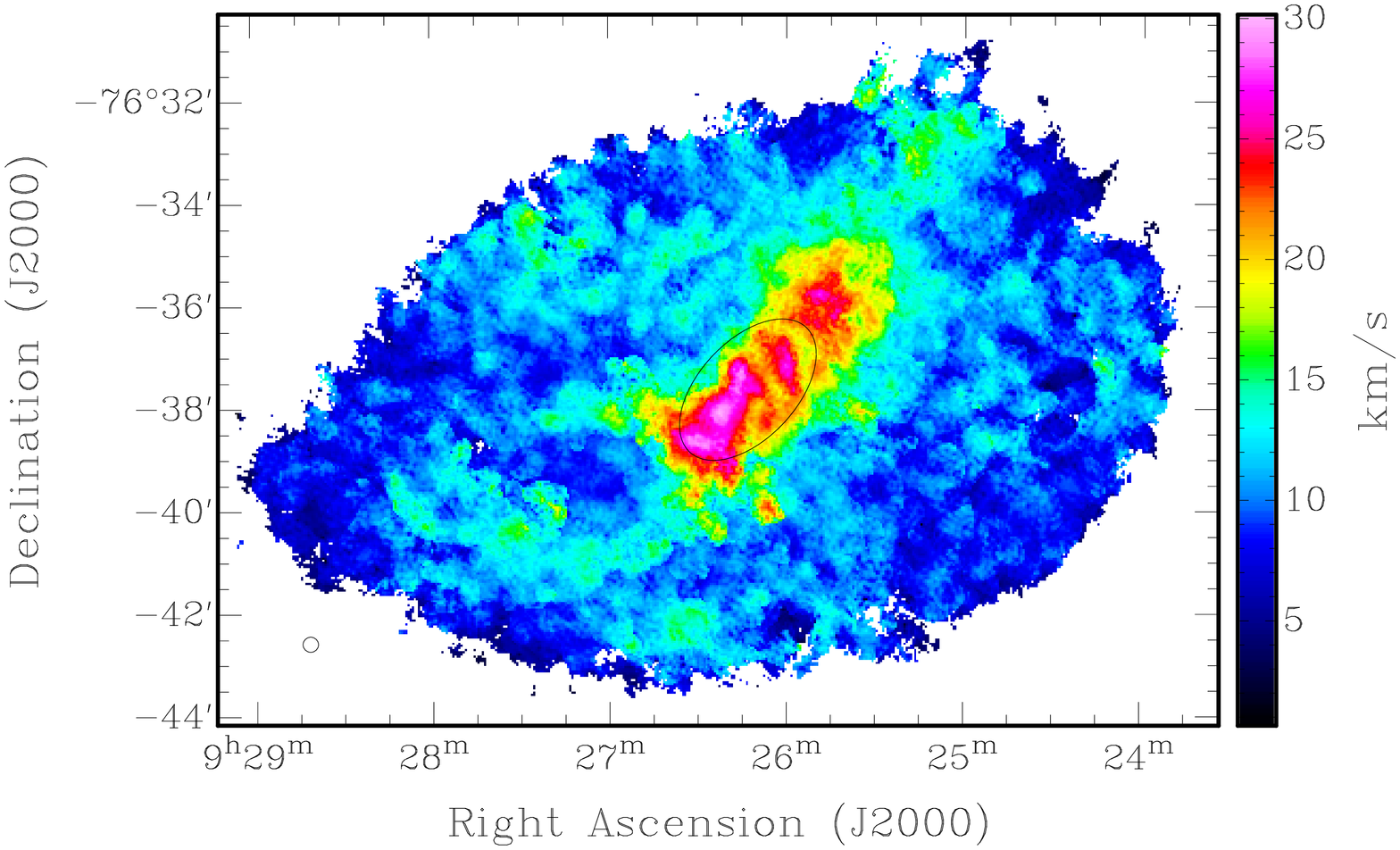}
	\caption{\textbf{Top left panel:} 3.6$\mu$m IRAC Spitzer image.  \textbf{Top right panel:} \hi total intensity map.  The two solid black lines represent the position-velocity slices shown in Fig. 2.  The open circle in the lower left corner represents the half power beam width.  \textbf{Bottom left panel:} \hi velocity field.  Contours are separated by 15 km s$^{-1}$ with the thick contour marking the systemic velocity at 465 km s$^{-1}$.  \textbf{Bottom right panel:} \hi velocity dispersion map.  In each panel the black ellipse represents the edge of the old stellar disk and has semi-major and semi-minor axes of lengths 100$''$ and  54$''$ (1.83 kpc and 0.9~kpc) respectively.}
	\label{moments}
	\end{centering}
\end{figure}

The lower right panel of Fig. \ref{moments} shows the \hi velocity dispersion map. The highest velocity dispersions are seen in the central-most regions of the galaxy where we observe ongoing active star formation and may therefore be due to the effects of stellar feedback from the young central stellar population.  The azimuthally averaged velocity dispersion profile yields an average outer disk dispersion of $\sim$ 10 km s$^{-1}$ which is close to the fixed gas velocity dispersion value of 11~km~s$^{-1}$ used by \citet{leroy_THINGS} for their sample of THINGS galaxies.  

\subsection{Non-circular motions}\label{NCM}
\begin{figure}[h]
	\begin{centering}
	\includegraphics[angle=-90,width=0.5\columnwidth]{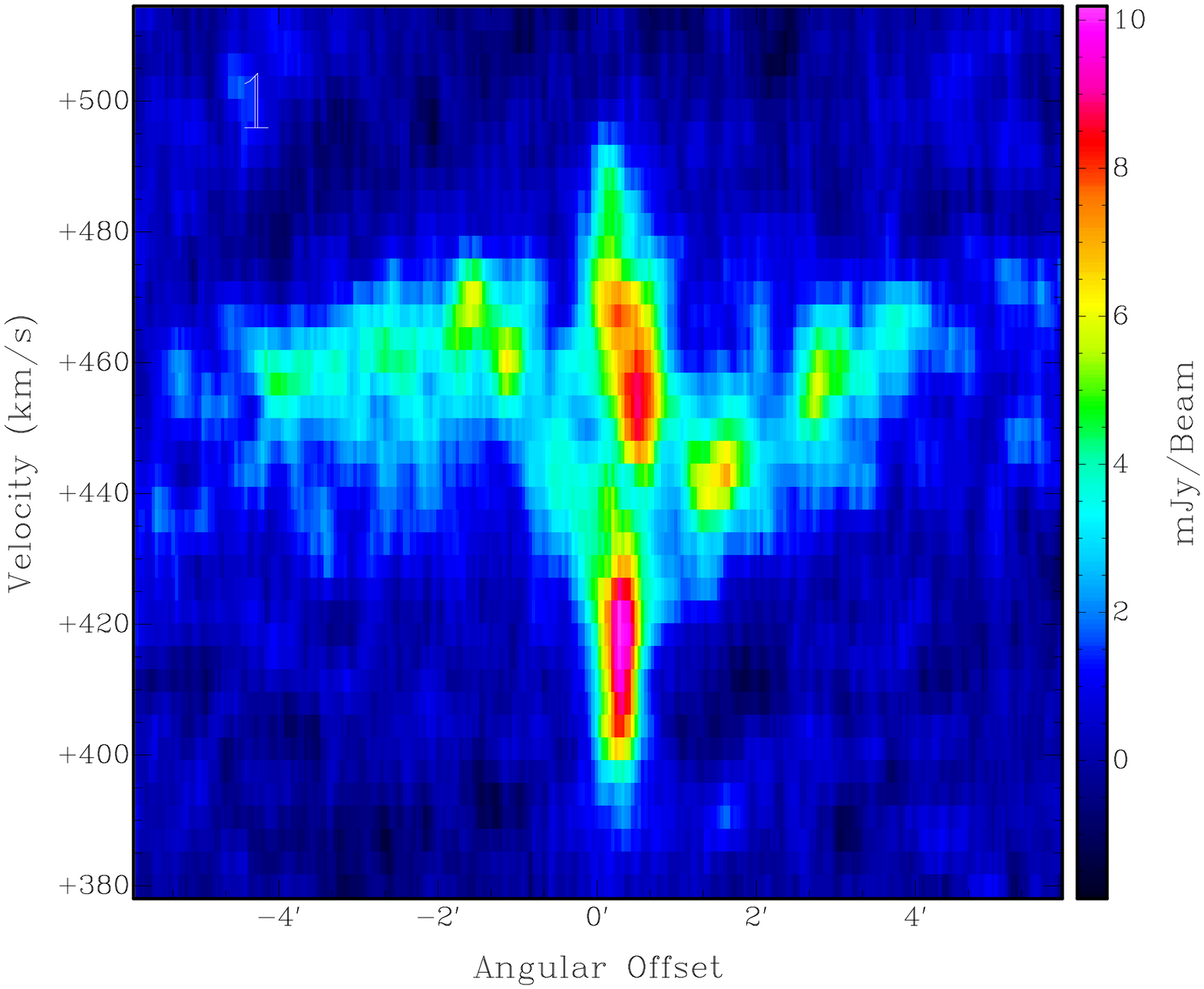}\includegraphics[angle=-90,width=0.5\columnwidth]{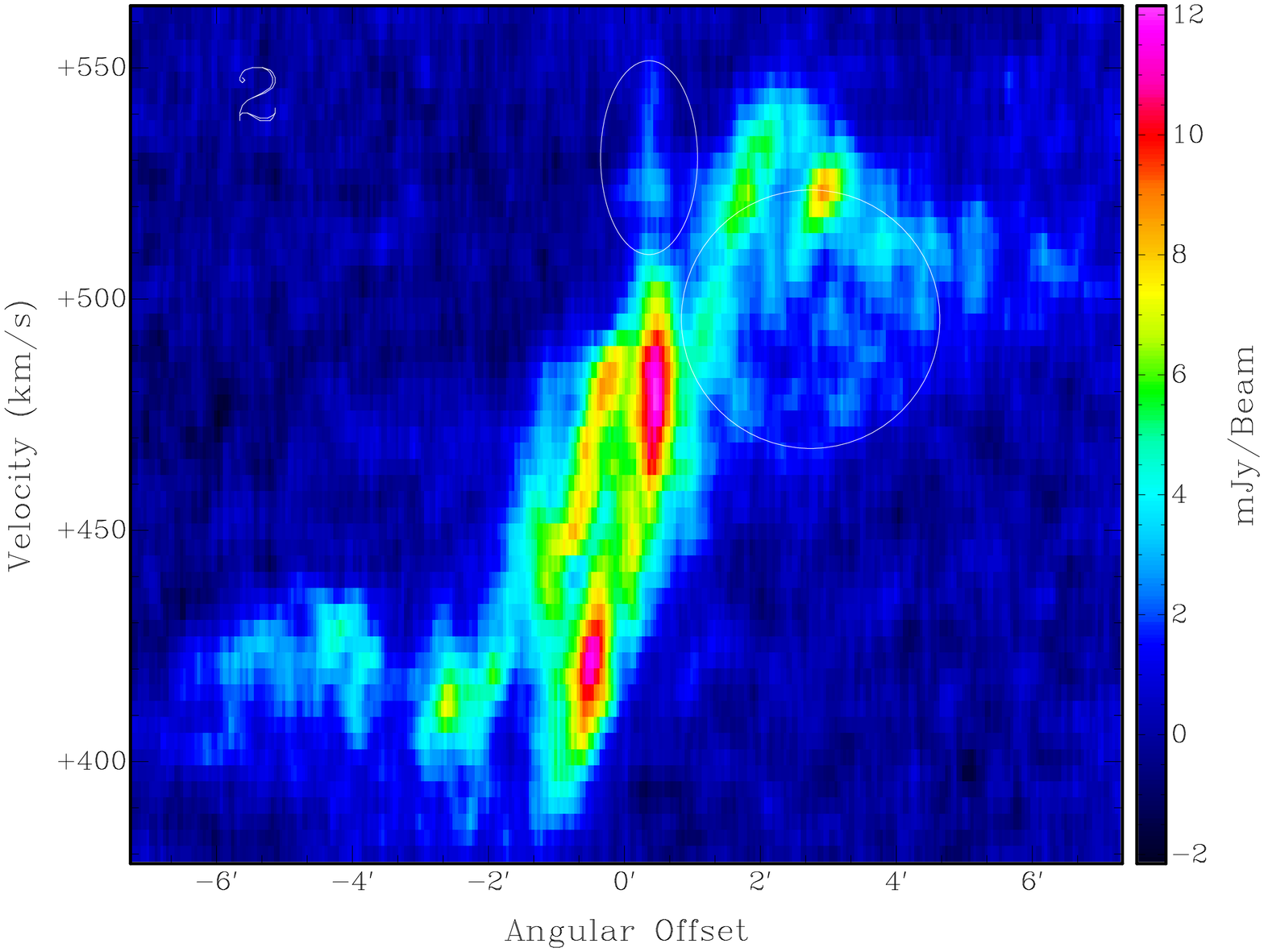}
	\caption{Position-velocity slices as represented by solid black lines in the \hi total intensity map.  The thickness of the slices is that of a single pixel in the data cube (2.5$''$) and the velocity resolution is \newline 3.5 km s$^{-1}$.  The white circle and ellipse in the right-hand panel enclose the \hi beard and velocity spike respectively.}
	\label{pv_slices}
	\end{centering}
\end{figure}

Position-velocity slices through the centre of NGC 2915 reveal the presence of non-circular gas motions (the slice positions and orientations are shown in the \hi total intensity map).  A possible fast-rotating gas component manifests itself as a sharp velocity spike, shown in the right-hand panel of Fig. \ref{pv_slices}.  The galaxy may also contain a central expanding gas component.  This is suggested by a large fraction of line profiles that are split by as much as $\sim 25$ km s$^{-1}$ (Fig. \ref{pv_slices}, left-hand panel).  Furthermore, line profiles near the centre of NGC 2915 are often skewed to one side.  Not surprisingly, these disturbed \hi line profiles are located within $\sim 150''$ of the centre of NGC 2915 thereby placing them within the immediate vicinity of the high-mass star-forming core.  It is thus likely that the injection of kinetic energy from massive stars into the inter-stellar medium is disrupting the gas dynamics.

\subsection{Gas infall}
The position-velocity slices reveal another interesting feature of the gas in the galaxy.  The receding half of the galaxy contains a gas component that is clearly lagging in velocity relative to the main gas disk of the galaxy.  This gas component appears as an \hi beard in the position-velocity slice (right panel, Fig. \ref{pv_slices}).  When this lagging gas is isolated it is found to coincide spatially with the plume-like \hi feature seen in the \hi total intensity map.   Since NGC 2915 is seemingly isolated, such a gas component may therefore provide evidence of gas being accreted from inter-galactic space onto the outer disk of the galaxy.  Indeed, gas accretion is directly observed in nearby systems and is thought to be the mechanism by which a galaxy's gas reservoir is replenished \citep{sancisi_review}, fuelling star formation in the inner disk.  A detailed metallicity analysis of NGC 2915 may therefore provide conclusive evidence for the infall of pristine gas.

\section{Central energetics}
\begin{figure}[h]
	\begin{centering}
	\includegraphics[angle=-90,width=0.6\columnwidth]{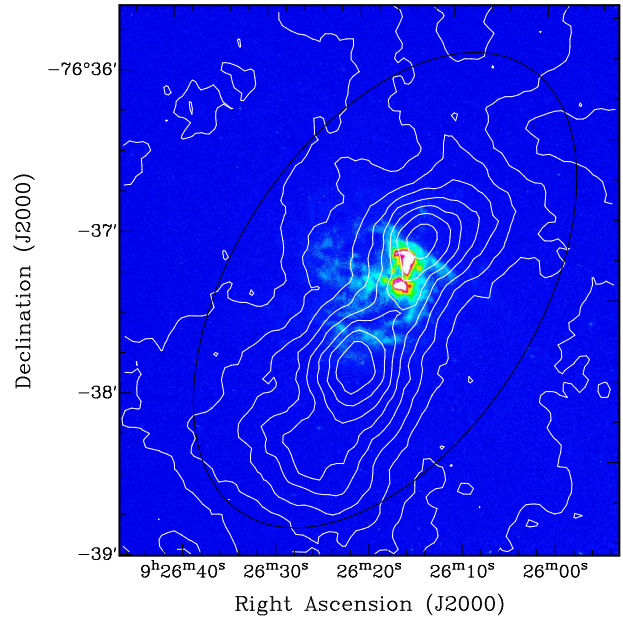}
	\caption{H$\alpha$ image from \citet{gil_de_paz_et_al_2003} with \hi total intensity contours overlaid.  \hi contour levels are in steps of 20~mJy~beam$^{-1}$.  The central-most contour of each \hi over-density is at a level of 200~mJy~beam$^{-1}$.  The black ellipse is the same one shown in the moment maps.}
	\label{halpha}
	\end{centering}
\end{figure}

H$\alpha$ emission \citep{gil_de_paz_et_al_2003} is observed near the centre of NGC2915, being contained within the two central \hi over-densities seen in the \hi total intensity map (Fig. \ref{halpha}).  The location of this H$\alpha$ emission also coincides with the observed split and broad \hi line profiles in the central region of NGC 2915.  This may therefore be evidence of a highly dispersed, expanding gas component.  We estimate the energy associated with this gas component to be $\sim 5\times 10^{46}$ J.   The H$\alpha$ luminosity of NGC 2915 is measured to be H$\alpha \sim 10^{32.5}$ J s$^{-1}$.  Such an energy output can account for the central gas energetics within $\sim5\times10^6$ years, which is the typical main sequence lifetime of a high-mass O or B type star.  We therefore argue that the stellar winds from the young stellar core dictate the inner gas dynamics of NGC 2915 and expel the gas outwards, thereby emptying the inner region of the galaxy.  Results from modeling of the central region as a rotating, expanding \hi component find that this simple dynamical model can account for most of the anomalous features seen in the \hi data products (Elson et al. in prep).

\acknowledgments
E. C. Elson would like to thank the South African Square Kilometre Array Project for generous funding throughout the duration of this PhD and  for the travel grant to attend the PRA meeting.  All authors wish to thank the conference organisers for an excellent meeting.


\end{document}